# Fusion Magnet Quench Risk Increase With Irradiation Damage

Jacob M. John, Mark R. Gilbert, Chris D. Hardie

*UKAEA, Abingdon, OXON, OX14 3DB, UK*

*Abstract*—Superconducting material enables fusion reactor magnet concepts to operate with current densities that would melt materials with non-zero resistance. The application of superconducting material is considered essential for net-positive power machines. Catastrophic damage can occur when superconductivity is lost and the current generates heat. This scenario is called a quench. Stabilizer material carries the magnet current (typically copper) during a quench and is the focus of this work.

Irradiation-induced defects store energy in the Cu crystalline lattice. The presence of defects reduces thermal conductivity (thermally insulating the superconductor), electrical conductivity (increasing temperature ramp rate during a quench), and specific heat capacity (increasing thermodynamic instability). The release of stored energy in the magnet materials, in combination with the magnet material property changes, has the potential to cause extreme off-normal events in superconducting magnets that worsen, at an increasing rate, with fluence.

Stored energy can be released causing local heating and increasing the risk of a quench. For example, following irradiation at 4.6K and an estimated fluence of $0.45 \times 10^{18}$ n/cm$^2$, an energy release of 0.023 J/g was measured from Cu when increased in temperature from 10K to 18K, which would have been enough energy to create the same temperature increase spontaneously.

Extrapolations of experimental data are used to estimate when spontaneous heating can occur due to the release of energy stored in irradiation-induced defects. Critical fluence values are estimated between $1.74 \times 10^{18}$ n/cm$^2$ and $2.85 \times 10^{19}$ n/cm$^2$ for neutron irradiation of Cu at a temperature of 20K.

High-temperature ramp rate in-situ cryogenic calorimetry experiments of magnet materials following irradiation would provide more clarity to designers of fusion magnet systems. Due to the increased quench risk with superconducting magnet dose, magnetic confinement reactor designers should consider the frequency of maintenance temperature cycles to maintain an appropriate level of risk during operation.

*Index Terms*—Fusion magnet, quench risk, superconductor, irradiation damage, stabilizer, neutron, resistivity change, Wigner energy.

## I. INTRODUCTION

WIGNER energy release was famously held responsible for the Windscale fire in 1957 [1]. The specific stored energy was greater than the specific heat capacity. The result was a self-heating system characterized by an effectively negative specific heat capacity. Work conducted at Oak Ridge National Laboratory demonstrated this mechanism in graphite [2].

In this work, the consequences of stored energy due to defect production, and the associated material property changes, are considered for superconducting magnets experiencing neutron irradiation. Superconducting magnets consist of superconductors, stabilizers, and structural materials. The stabilizer

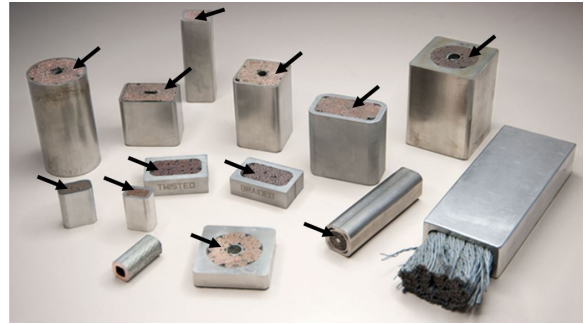

Fig. 1: HTS magnet cross-section contains plenty of Cu. Cu is shown with arrows. This work considers how energy stored in this material due to irradiation damage to the crystalline lattice affects the operation of the machine. Picture repurposed from Salazar et al. [5].

material portion of the magnet is primary the focus of this work.

The critical current and critical temperature of superconductors degrade with neutron irradiation [3], [4]. Critical current describes the current density capacity and critical temperature describes the transition temperature between superconducting and normal conducting state. The uncontrolled transition from superconducting to a normal conducting state during operation is called a quench. The current in the magnet is preferentially carried by the stabilizer material during a quench, and as a result, generates heat. Typically, the stabilizer material is Cu. Figure 1 shows the large amount of Cu in a magnet conductor cross-section. The stabilizer material is designed for strain relief, provides a thermal pathway to cool the superconductor, and must carry current during a quench. As such, changes to the material properties of the stabilizer have a significant impact on the performance of the magnet.

Irradiation of the stabilizer material's crystalline lattice displaces atoms, producing point defects and defect structures. The defects present obstacles to the motion of electrons, reducing the thermal and electrical conductivity of the material. The Wiedemann-Franz Law [6] presents a linear relationship between thermal and electrical conductivity due to the near-linear dependence of both of these properties on the mobility of electrons for Cu. The defects store strain energy in the lattice, which is released with recombination and annihilation. Defects become mobile at elevated temperatures and the energy released can be interpreted as a reduction of specific heat capacity as observed by Blewitt et al. [7]. Upon recombination,

TABLE I: Properties of radiation-induced point defects [8].

| | Cu [eV] | Al [eV] |
|---|---|---|
| Interstitial Formation Energy | 2.2 | 3.2 |
| Vacancy Formation Energy | 1.27 | 0.66 |
| Interstitial Migration barrier | 0.12 | 0.12 |
| Vacancy Migration barrier | 0.8 | 0.62 |

the loss of thermal and electrical conductivity is recovered proportionally.

During the formation of a Frenkel pair in Cu, 3.5 eV of the required 20 eV displacement energy is stored in the lattice, Table I shows formation energies for point defects in both Cu and Al. If the potential energy of a single Frenkel pair in Cu were to transform into thermal vibrations, the Boltzmann factor described by the equipartition of energy theory can be used to estimate that a volume of 2030 atoms could be heated from 20K to 40K. Consider a case where 1 in every 2030 atoms is displaced as a Frenkel defect. The material in this case has experienced a dose of 0.00049 displacements per atom (DPA). This dose prediction is low relative to the critical current degradation rate of HTS. Iliffe et al. observed that the critical current of REBCO HTS was measured to degrade to zero following proton irradiation to a dose of 4 mDPA at 40K [4]. Dependent on accumulated dose through service, the storing of Wigner energy in the Cu stabilizer of a superconducting fusion magnet may be a necessary consideration in the design and operation of a magnetic confinement fusion reactor. This estimate is accurate if all displacements in the Cu are Frenkel pairs, but in reality defect structures would form during irradiation to reduce the free energy in the lattice. To predict the dose at which a self-heating system is possible from 20K to 40K, it is necessary to account for the effect of representative defect structures.

In this work, the effect of defect structures is accounted for by referencing experimental measurements of defect recovery dynamics as temperature increases following cryogenic irradiation. Experimental data from literature are leveraged to predict heat release profiles due to defect recombination for different irradiation scenarios. Saturation phenomena are considered to provide bounds for the applicability of these predictions. The heat release profiles are used to predict the reducing temperature margin for the fusion magnet, that is between operation and quench runaway, as a function of neutron dose. This definition of temperature margin is new and accounts for defect energy release and reduction in the high-temperature superconductor's critical temperature. The process of producing these predictions leads to the definition of two new engineering parameters that characterize the fusion magnet's risk against quenching: critical span and the Well's number.

To date experiments investigating the self-heating of irradiated cryogenic Cu have not been undertaken, which is likely due to the expense/difficulty in producing these conditions, as well as a perceived lack of importance. However, with the need to build superconducting magnets for fusion reactors, this mechanism and its potential impact on power plant operation must be understood.

## II. RESISTIVITY CHANGE

Resistivity increases as defect density increases. Losehand et al. [9] reported an increase of 0.27 n$\Omega$m in Cu due to neutron irradiation to an unreported fluence at 4.6K. The measured relationship between fluence and resistivity change reported in Nakagawa et al. [10] (up to a fluence of $1 \times 10^{19}$ n/cm$^2$) is used to estimate the fluence of the Losehand et al. experiment. Nakagawa et al. report a fluence of $4.5 \times 10^{17}$ n/cm$^2$ upon measuring a resistivity change of 0.27 n$\Omega$m in Cu at 4.6K following neutron irradiation. Therefore, the Losehand et al. experiment is taken to have performed energy storage measurements at a fluence of $4.5 \times 10^{17}$ n/cm$^2$ in this work. The application of the Nakagawa relationship to the Losehand results here is considered a valid assumption as both experiments took place in the same Munich Research Reactor and the resistivity measurements were performed at the same laboratory (about 8 years apart).

The stabilizer section of magnets should have as low resistivity as possible due to heat generation during a quench. Typically, high-purity Cu with a high residual resistance ratio (RRR) is applied in design. RRR describes a ratio of resistivity at room temperature to that measured at 4.6K, $\rho_{300K}/\rho_{4.6K}$. RRR is closely linked to purity since it is impurities that determine the resistivity at 4.6K. Impurities from nuclear transmutation, radiation-induced defects, or dislocations from local yielding can reduce RRR. To manufacture and install a fusion magnet while maintaining a RRR of 100 in the stabilizer material is a feat of engineering, since handling the material can significantly reduce the value. 100 RRR Cu is measured to have a resistivity of 0.17 n$\Omega$m at 20K, so an increase of 0.27 n$\Omega$m following a fluence of $4.5 \times 10^{17}$ n/cm$^2$ would have a substantial effect on the performance of the magnet during a quench.

A lower initial resistivity of a material corresponds to a greater increase in RRR following irradiation. For example, the resistivity of 1490 RRR Cu was reported to increase by 3.21 n$\Omega$m following irradiation to a fluence of $8.49 \times 10^{18}$ n/cm$^2$. This represents a 267.5 multiplication from the initially measured resistivity of 0.012 n$\Omega$m [11]. Using a high RRR material for the stabilizer material does not necessarily result in increased magnet performance during a quench due to the high rate of resistivity change with fluence. Resistivity increases reported from many different cryogenic Cu neutron irradiation experiments are documented in Table II and used in further analysis in this work.

## III. STORED ENERGY RELEASE

Energy capacity and release rate are key metrics for energy storage devices. Typically, if the specific capacity is high this comes at the cost of the specific release rate. Compared to typical energy storage mechanisms, the specific capacity of Wigner energy in cryogenic Cu is low, but the release rate is high. Similarly, the capacity of capacitors is low but the energy release rate is high, compared to batteries. The energy release kinematics of a commercial 21700 type battery cell, a commercial Skelcap SCA3200 supercapacitor, and cryogenically irradiated pure Cu are compared in Figures

TABLE II: Resistivity change of Cu due to cryogenic neutron irradiation [12].

| Temperature K | RRR $\rho_{300K}/\rho_{4K}$ | Before nΩm | After nΩm | Fluence $10^{18}/cm^2$ | Spectrum Description | Change per neutron $[\Delta n\Omega m]/[10^{18}/cm^2]$ | Study Reference |
|---|---|---|---|---|---|---|---|
| 4.4 | Not Reported | 0.016 | 0.026 | 0.012 | fission | 0.84 | 1962 Coltman [13] |
| 4.8 | 5400 | 0.0031 | 0.0039 | 0.029 | thermal | 0.02 | 1967 Coltman [14] |
| 4.6 | 118 | 0.14 | 0.16 | 0.036 | fission | 0.56 | 1967 Burger [15] |
| 4.6 | 453 | 0.037 | 0.056 | 0.036 | fission | 0.52 | 1967 Burger [15] |
| 4.6 | 519 | 0.033 | 0.93 | 1.18 | >0.1 MeV | 0.76 | 1970 Böning [12] |
| 4.6 | 16.97 | 0.033 | 0.93 | 1.18 | >0.1 MeV | 0.76 | 1970 Böning [12] |
| 4.6 | 714 | 0.024 | 0.90 | 1.18 | >0.1 MeV | 0.74 | 1970 Böning [12] |
| 4.5 | 2280 | 0.0082 | 1.17 | 2.0 | >0.1 MeV | 0.58 | 1972 Horak and Blewitt [16] |
| 18 | 492 | 0.038 | 0.96 | 1.23 | >0.1 MeV | 0.75 | 1974 Brown [12] |
| 4.5 | 882 | 0.019 | 0.024 | 0.57 | thermal | 0.01 | 1975 Horak and Blewitt [17] |
| 4.5 | 987 | 0.017 | 0.41 | 0.58 | >0.1 MeV | 0.67 | 1975 Horak and Blewitt [17] |
| 4.5 | 2280 | 0.0080 | 1.16 | 2.0 | >0.1 MeV | 0.58 | 1975 Horak and Blewitt [17] |
| 4.6 | 432/1490 anneal | 0.012 | 3.23 | 8.49 | >0.1 MeV | 0.38 | 1977 Nakagawa [11] |
| 4.6 | 432/1464 anneal | 0.012 | 3.23 | 8.49 | >0.1 MeV | 0.38 | 1977 Nakagawa [11] |
| 4.2 | 1432 | 0.012 | 0.021 | 0.0045 | 15 MeV d-Be-neutrons | 2.11 | 1977 Roberto [18] |
| 5 | 950 | 0.018 | 1.36 | 2.52 | >0.1 MeV | 0.53 | 1970 Böning [12] |
| 4 | 2180 | 0.0078 | 0.058 | 1.8 | thermal | 0.03 | 1979 Williams [19] |
| 5 | 900 | 0.015 | 0.044 | 0.042 | fission | 0.69 | 1981 Takamura and Kato [20] |
| 5 | 150 | 0.092 | 0.12 | 0.042 | fission | 0.76 | 1981 Takamura and Kato [20] |
| 4.2 | 250 | 0.083 | 0.24 | 0.073 | 14.8 MeV RTNS | 2.22 | 1983 Van Konynenburg [21] |
| 4.2 | 250 | 0.095 | 0.27 | 0.084 | 14.8 MeV RTNS | 2.15 | 1983 Van Konynenburg [21] |

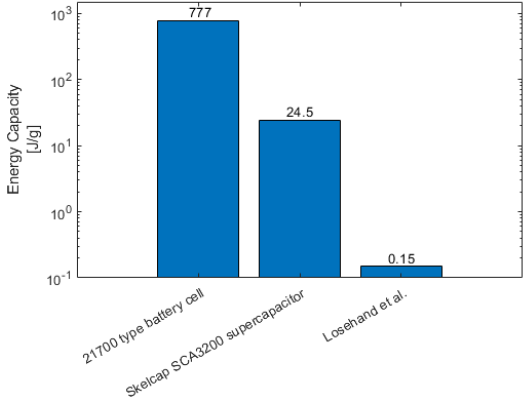

Fig. 2: Energy storage mechanisms balance energy capacity and power capacity. If the energy capacity of a system is very high, then it is likely that the release of the stored energy will happen slowly.

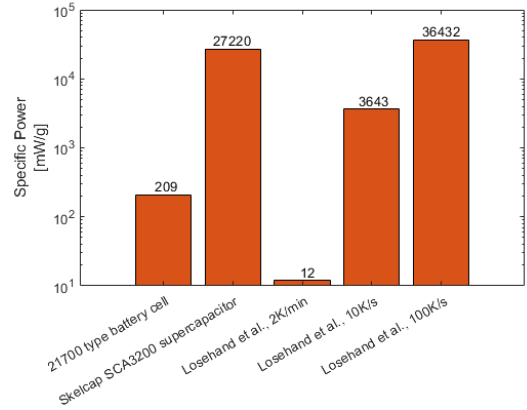

Fig. 3: With reducing energy capacity, power capacity increases. Stored energy release due to interstitial defect annealing is presumed to showcase a power capacity greater than that of capacitors.

2 and 3 to provide an intuitive and practical understanding of their energy storage performance.

The room temperature performance of a 21700 type battery cell [22] and SkelCap SCA3200 supercapacitor [23] are compared against the energy storage and release performance of Cu irradiated at cryogenic temperature reported in the Losehand et al. [9] experiment. The researchers measured energy released by the Cu while heating from 5K to 300K at 2K/min [9]. For this comparison of energy storage and release performance, we are only concerned with the energy release that occurred between 20K to 40K, as this portion of the energy release is most relevant against the operation conditions of superconducting fusion magnets.

During the Losehand et al. [9] experiment, temperature was increased slowly to ensure complete annealing of defects. Figure 4 shows the heat released and resistivity recovered due to annealing during the 2K/min heating following cryogenic neutron irradiation. If heating is too fast, defects do not have enough time to migrate and annihilate before more stable defects become mobile, distorting the measured heat release profile.

The annealing rate is either limited by temperature ramp rate or defect mobility. Material property dynamics can be interpreted using methods developed for system dynamics control in the field of modern control theory[1], this perspective may prove useful for modeling the dynamics of material properties during experiments or operations.

In this work, we hypothesize that the annealing rate of interstitial defects (stage I recovery) in 20K pure Cu scales linearly with increases in temperature ramp rate up to 71K/s, and potentially higher. Literature has not yet experimentally determined the temperature ramp rate at which the linear scaling of stage I annealing rate in Cu breaks down. Interstitial defects migrate and annihilate more readily than defect structures due to lower threshold energy for mobility[2], so the linear scaling temperature ramp rate limits are different for different stages of recovery. Temperature ramp rates above 71K/s are not expected during a quench and are of less interest to this work. The hypothesized upper limit for temperature ramp rate linear scaling is derived from a comparison to a more standard energy storage and release mechanism. It is assumed that the specific power of the stage I recovery in Cu scales linearly with temperature ramp rate up to, and at least, the specific power of a SkelCap SCA3200 supercapacitor, once the energy release is scaled by the ratio of specific heat capacity to account for the temperature differences of the mechanics[3], implying that the linear scaling of specific power for Stage I annealing in Cu is valid up to, at least, 71K/s. Figure 3 showcases specific power predictions comparing temperature ramp rates that are relevant for fusion magnet quench systems, 10K/s and 100K/s. The method used here to predict the temperature ramp rate limit of the linear scaling of specific power is practical, however, the sophisticated molecular dynamic question is reduced using assumptions produced from an intuitive comparison of different technologies. The implications of this prediction being wrong as a distortion in the heat release profile during annealing and distorted high-level system dynamics predictions in the analysis section of this work. For this work, this method provides a suitable level of precision, however, detailed molecular dynamics studies on the relationship between temperature ramp rate and the profile of Stage I annealing would be a necessary step to increase the confidence of these predictions.

---

[1] The temperature ramp rates of 2K/min used by Losehand et al. [9] are negligible relative to defect recombination time constants. The time constant is a parameter of linear time-invariant dynamic systems. The time constant is the period until the state(s) describing a dynamic system exhibit 63% of the total change resulting from a step input. The temperature ramp rate is low enough relative to defect mobility to characterize the sample in this study as quasi-isothermal.

The analysis in this work interprets the relationship between dose, temperature, and defect recovery using the recovery curve presented in Losehand et al. [9]. The curve presented is conducted at a single dose/fluence. To preserve the temperature variation of the observation, the curve was divided by its integral. The curve was also divided by the predicted fluence of the measurement. The assumed linear relationship between total stored energy and dose (up to saturation, and explored in the next section) is used in the analysis section of this work by multiplying the processed curve by fluence and total stored energy of other experimental observations consisting of single dose/fluence measurements. This approximation enables analysis at various irradiation conditions without performing a difficult set of experiments.

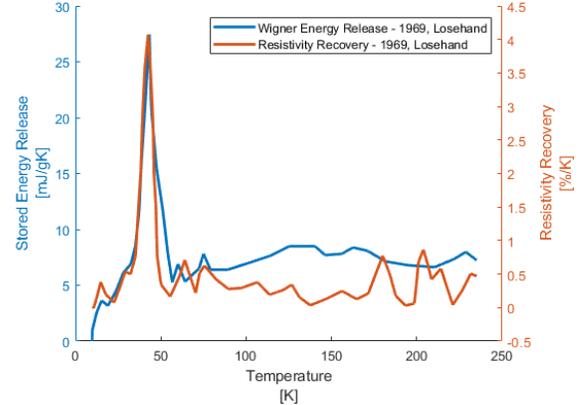

Fig. 4: Defect recovery peak occurring at 45K. Most fusion HTS magnets quench around 40K, suggesting the increased likelihood of a quench with irradiation. Wigner energy release (blue curve, left axis) and normalized resistivity recovery (red curve, right axis) of cryogenic Cu post neutron irradiation at 4.6K inside the Munich Research Reactor. Experiment performed by Losehand et al. and replotted here [9].

The remaining number of Frenkel pairs following a collision is a function of many factors. The conditions of cryogenic neutron irradiation are favorable for the survival of a substantial number of Frenkel pairs. The low temperature of the lattice reduces rates of migration and so reduces the frequency of defect annealing. It is this residual number of Frenkel pair defects in the magnet's stabilizer material that primarily presents a risk to the operation of superconducting fusion magnets. The defect microstructures relevant to conditions anticipated for a typical reactor are discussed in the following section.

## IV. Defect Saturation

The Saturated defect density of a target increases with decreasing temperature due to reductions in the likelihood of interstitial and vacancy defect interaction. Primary knock-on collisions during irradiation produce defects and after the transient stages of a collision cascade (where significant annealing occurs), point defects and defect structures remain in the lattice. Interstitial defects migrate more readily than vacancy-type defects. Table I shows migration and formation energies for point defects in Cu and Al. These materials are often discussed in conceptual applications for superconducting fusion magnets, this table provides a comparison of the behavior of point defects for both materials.

Both first-order and non-first-order recovery have been observed during Stage I recovery of copper, this implies different mechanics are responsible for migration and annihilation at different substages of Stage I recovery [30]. Short-range or long-range recovery processes will dictate the rate of annealing at different substages, so the remaining defect microstructure after a collision cascade will play a key role in the dynamics of the annealing. For long-range recovery substages, interstitial defects probabilistically favor migration along stress gradients in the lattice [31].

---

[2] Point defects and complex defect structures form during the transient stages of a collision cascade. Frenkel pairs agglomerate to form defect structures to reduce free energy in the lattice. Not all Frenkel pairs form defect structures. Irradiation experiments of Cu at temperatures ranging from 5K to 17K show show that Frenkel pairs anneal at temperatures lower than 45K and complex defect structures anneal at temperatures greater than 45K [7], [9], [24]. The temperature of the Frenkle pair peak annealing rate does not change with dose or between experiments, however, local peaks above 45K change shape with dose due to varying defect structures [25]. Because interstitial defects are more mobile than complex defect structures, Stage I annealing (occuring below 45K) can occur at a greater rate than higher-order stages of annealing.

[3] It is noteworthy that the heat load of interstitial defect annealing occurs at cryogenic temperature. The specific heat capacity of pure Cu is 8 J/kgK and 385 J/kgK at 20K and 300K, respectively [26], representing an increase by a factor of approximately 48. Cu only requires 0.5812 J/g to heat from 20K to 40K. To compare the specific power of interstitial defect annealing against that of a supercapacitor in Figure 3, we multiply by 48.

TABLE III: Wigner energy release of Cu due to cryogenic irradiation from different spectra.

| Temperature K | Particle | Fluence $10^{18}$/cm$^2$ | E. Released J/g | E. Released per particle [J/g]/[$10^{18}$/cm$^2$] | Study Reference |
|---|---|---|---|---|---|
| 17 | X-10 Neutron | 1.2 | 0.04 | 0.033 | 1959 Blewitt [7] |
| 5 | 1.2MeV Electron | 0.9 | 0.1 | 0.11 | 1959 Meechan [27] |
| 15 | 11 MeV Deuteron | 0.0029 | 0.04 | 13.79 | 1960 Granato [28] |
| 15 | 11 MeV Deuteron | 0.0083 | 0.09 | 10.84 | 1960 Granato [28] |
| 5 | FRM I (Munich Research Reactor) Neutron | 0.45 (Section II) | 1.74 | 3.87 | 1969 Losehand [9] |
| 5 | HFIR Neutron | 0.13 | 0.03 | 0.23 | 1990 Richard [29] |
| 5 | HFIR Neutron | 1.24 | 0.4 | 0.32 | 1990 Richard [29] |
| 5 | HFIR Neutron | 3.5 | 0.84 | 0.24 | 1990 Richard [29] |

Upon coalescence, the defects annihilate and release stored energy as heat. The rate of defect production and annihilation are equivalent at the defect concentration saturation value, so any mechanism that affects migration rates will change the expected defect concentration value. It was shown recently that defect migration rates can increase when materials are exposed to a gamma flux [32], complicating defect saturation concentration predictions for materials experiencing cryogenic neutron and gamma irradiation. Factors such as temperature, spectrum energy, material, lattice stress, and fluence affect the defect production and annihilation rates, and so the defect saturation value.

A simple model describes a "spontaneous recombination volume" to predict defect concentration saturation [33]. If an interstitial and a vacancy both exist inside this volume, the two annihilate. The radius of this decreases with decreases in temperature, and so the maximum number of point defects that can coexist in a 3-dimensional volume increases to the third power with reductions in temperature. In other words, opposing defect types are more likely to annihilate with temperature increases. Defect concentration saturation is predicted at 0.16 at% for cryogenic FCC metals, equivalent to 2.509 mDPA [24]. This equates to one Frenkel pair for every 625 atoms; thus, when considering a generic FCC metal at cryogenic temperature, continued irradiation would result in Frenkel pair recombination at the same rate as production. The behaviour of defects produced by irradiation in some FCC metals, such as Au [11], deviates significantly from other materials, so results from this model prediction must be generalized between FCC materials with discretion and only between materials that are shown to exhibit similar recovery dynamics. Defect concentration deduced from resistivity measurements in Ni-0.1 at% $^{235}$U was found to saturate at 0.14% at 7K [34], confirming model predictions for Ni. Cu and Ni are shown to have similar defect recovery behaviour [11], so this saturation limit model is considered valid for both metals.

The maximum possible specific stored energy is predicted by assuming all defects are Frenkel pairs. The stored energy in this case can be easily calculated by multiplying the energy contained by each Frenkel defect by the number of Frenkel defects,

$$E_{\text{stored}} = \frac{DPA \times N_A \times E_{\text{FP}}}{M} \quad (1)$$

Where DPA is dose at saturation (2.509 mDPA), M is molar mass of Cu (63.5 g/mol), $N_A$ is the Avogadro constant ($6.02 \times 10^{23}$ Atom/mol), and $E_{\text{FP}}$ is the sum of the formation energy of both interstitial and vacancy defects ($5.6 \times 10^{-19}$ J). The maximum possible stored energy, $E_{\text{stored}}$, for this system is 13.3 J/g. The presence of defect structures would reduce the specific stored energy in the lattice. This estimate is compared against experimental measurements to interpret if the measurements are physically realistic.

Richard et al. [29] suggest the onset of a Wigner energy saturation trend in cryogenic Cu. Three data points at 0.031 J/g, 0.41 J/g, and 0.85 J/g in Figure 5 seem to suggest a saturation trend taking place with the highest fluence reported at $3.5 \times 10^{18}$ n/cm$^2$. Outliers are present above and below the curve drawn between these data points. Also, all measured values are below the maximum physically feasible specific energy storage limit of 13.3 J/g predicted using equation (1). More experimental data points are required for resistivity change and stored energy due to irradiation from neutrons for fluences above $2 \times 10^{18}$ n/cm$^2$ at cryogenic temperatures.

All of the energy release values presented in the fourth column of Table III were produced by Differential Heat Flux Calorimetry, using thermocouples, of an irradiated sample and a sample at the control state. The sample at control state was either a non-irradiated sample with temperature measurement occurring simultaneous to irradiated sample measurement during heating, as is possible for the experiments performed at accelerator facilities, or the control state measurement was provided by comparing temperature versus time profiles across heating cycles of the same sample. The slope of the temperature versus time measurements provides a measurement of specific heat capacity, and comparing specific heat capacity between signal and control measurements provides an energy release signal due to defect annealing.

The orange dots in Figure 5 show a collation of experimental measurements for stored energy release following cryogenic irradiation from neutron, deuteron, and electron fluxes. The scatter between the dots is evidence of the sensitivity of the energy storage mechanism to changes in irradiation conditions. The gradient of each data point is shown in Figure 6, further describing that the rate of change of energy storage and resistivity increase during irradiation is a strong function of irradiation conditions and that none of the data points can be omitted as unrealistic outliers.

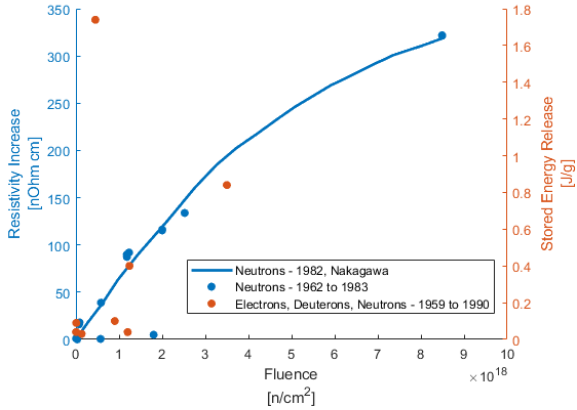

Fig. 5: Results from various reported irradiation experiments show an apparent correlation between resistivity increase (left axis, blue dots/line, Table II) and stored energy (right axis, orange dots, Table III) in cryogenic Cu irradiations.

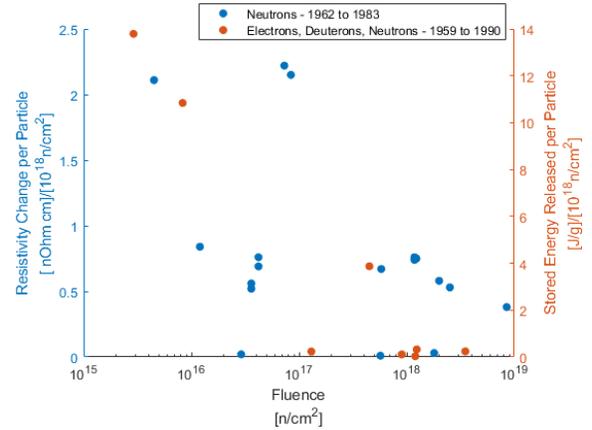

Fig. 6: Dividing the magnitude change for each data point in Figure 5 by the corresponding fluence shows the gradient of each data point, showing that stored energy is a strong function of irradiation conditions and communicating why new experiments need to be performed.

The Losehand et al. data point in Figure 5 at $0.45 \times 10^{18}$ n/cm² and 1.74 J/g stands out from the rest, but is not an unrealistic measurement. The Losehand et al. gradient data point at $0.45 \times 10^{18}$ n/cm² and 3.87 [J/g]/[$10^{18}$/cm²] is higher than the other measurements produced following neutron irradiation and this deviation may be due to differences in spectra energy leading to differences in primary collision energy. The deuteron irradiation shows much higher energy stored per particle, due to higher primary knock-on energy and increased likelihood of interaction, due to charge, leading to more collisions per particle and thus more stored energy. In the analysis of this work, the Losehand et al. data point is taken as the upper limit for energy storage, while the Richard et al. data points are interpreted to provide a less extreme comparison, but not necessarily more accurate, prediction of energy storage in the Cu lattice due to cryogenic neutron irradiation.

## V. ANALYSIS

### A. Quench Case Study

Superconducting fusion magnets carry current densities that would melt materials with non-zero resistance. Temperature fluctuations, for any reason, present a threat to the stable operation of the magnet. Loss of superconductivity during operation is called a quench. Quenches are characterized as a rapid release of stored energy. In a quench scenario, current shifts from the superconductor to the stabilizer material due to loss of superconductivity. The stabilizer material is typically Cu and generates heat according to Joule's first law when carrying current. Joule heating is the dominant heat source during a quench. There are many factors to consider in the initiation, dynamics, and control of these events characterized by a rapid high energy density release.

Salazar et al. experimentally examined quench temperature dynamics [5]. VIPER cables were operated at 10K and a heater was used to provide a disturbance. Figure 7 shows the temperature response of the cable following 36 J, 45 J, 54 J, and 63 J heat inputs. The cable quenched on the fourth trial. Temperature dynamics following the 63 J disturbance are shown in figure 8. These figures were published in other works and are presented in this study to provide a realistic starting point for the interpretation presented in this work.

The results produced by Salazar et al. [5] and reported in Figure 8 clearly characterize the dynamics of a quench. The 63 J heat input resulted in a 6.4K/s ramp rate from 10K to 18K [5] during the heater pulse. The heater temperature and local voltage initially dropped after the end of the pulse, then continued to increase with an accelerating rate once the superconductor lost superconductivity. This stage of accelerating the increase of voltage and temperature characterizes quench state dynamics. The effect of neutron-irradiated stabilizer material on quench initiation can be estimated from Figure 8 with a few assumptions:

1) Temperature increase was entirely within the Cu.
2) Temperature measurement represents bulk temperature.
3) For the applied heating of 6.4K/s, the annealing rate is limited by the migration rate and not the heating rate (as explored in section III).
4) Stored energy release dynamics can be approximated by linear scaling with respect to total stored energy release from the Losehand et. al [9] results, Figure 4.

The 62 J heat input was enough to cause the cable to quench, Figure 8. 3000 g of Cu is predicted to have been heated. This was found by dividing 62 J by the amount of energy required to heat Cu from 10K to 18K (0.021 J/g). If irradiated to Losehand et al. [9] conditions, the Cu would have released an additional 0.023 J/g due to defect recombination. For this mass, this energy release corresponds to an additional 68.3 J of thermal energy from defect recovery, significantly increasing the risk of quench initiation. If the Cu were irradiated to the conditions presented in Richard et al. [29], 33.0 J would have been released due to defect recovery, still presenting a significant increase in the likelihood of quench initiation.

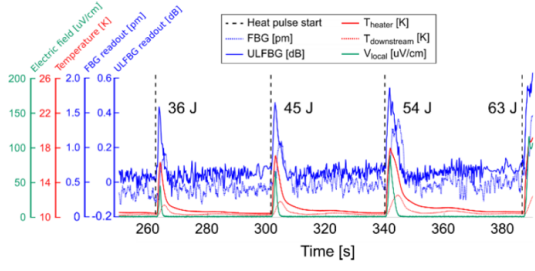

Fig. 7: VIPER cable signal output while carrying 41 kA and subject to an increasing thermal energy disturbance [5]. 36 J, 46 J, 54 J trials did not result in a quench whereas the 63 J trial did.

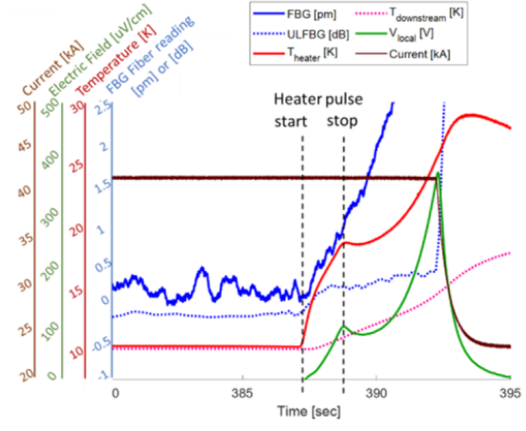

Fig. 8: 63 J thermal disturbance initiating a quench [5]. Close-up of the fourth trial in Figure 7. See the rapidly increasing temperature signal given by the red line.

For the conditions considered and the assumptions applied, the predicted energy release due to defect recombination was equivalent, or greater than, the input energy disturbance that caused the cable to quench. This implies that it is highly likely that defect recombination would lead to an early onset of quenches and that the mechanism should be investigated further.

### B. Temperature Margin

The temperature margin is understood by industry as the difference between the operating temperature, $T_{operation}$, of the magnet and the temperature at which quench initiates, $T_{critical}$,

$$T_{margin} = T_{critical} - T_{operation} \qquad (2)$$

This work describes that the temperature margin reduces as the superconducting magnet is irradiated. To derive an expression that describes the dynamic behaviour of temperature margin with time (or fluence) we start by imagining the energy balance for an irradiated magnet, in accordance with Newton's Second law, heating from operational temperature to critical temperature,

$$W_{T_{operation}}^{T_{critical}} = C_{T_{operation}}^{T_{critical}} \qquad (3)$$

where $W_{T_{operation}}^{T_{critical}}$ is the energy released in the magnet by defect recombination during heating from operational temperature to critical temperature and $C_{T_{operation}}^{T_{critical}}$ is the energy required to heat the bulk of the magnet from operational temperature to critical temperature. The primary heating scenario of concern is when the magnet experiences an external energy input resulting in a temperature fluctuation,

$$W_{T_{operation}}^{T_{critical}} = \cancel{C_{T_{operation}}^{T_{disturbance}}} + C_{T_{disturbance}}^{T_{critical}} \qquad (4)$$

$C_{T_{operation}}^{T_{disturbance}}$ is the external energy input that initiates the disruption to stable operation, and $C_{T_{disturbance}}^{T_{critical}}$ is the remaining energy capacity between the resulting temperature increase and the critical temperature of the magnet. The remaining terms in equation (4) show how at some magnitude of external energy input, the energy released by defect recombination will be sufficient to initiate a quench. As defects accrue in the magnet materials, more stored energy is released upon an energy disturbance, and the tolerance to external energy input reduces. This is understood as a reduction in temperature margin.

The stabilizer material is in direct thermal contact with both the superconductor and the cryogenic coolant, it provides the only thermal pathway for cooling the superconductor, and it consists of orders of magnitude more mass than the superconducting layer in the magnet (ReBCO HTS layers are nominally 1 μm to 2 μm thick). Further, the stabilizer carries current during a quench. So understanding how the properties of stabilizer materials affect the temperature dynamics of the magnet is a must. Since stabilizer materials provide the medium of cooling, consist of a higher fraction of the magnet, and dictate the rate of heating during a quench, this energy release analysis will be performed for energy release due to defect recombination inside the stabilizer and not the superconductor. Superconductor performance degradation can be accounted for by examining reductions in critical temperature with fluence [3].

Temperature margin is the maximum temperature disturbance that the magnet can experience before a quench initiates, so when predicting temperature margin the terms denoted by "disturbance" in equation (4) are replaced with denotation of "disturbance". To describe temperature margin, in the general sense, as the magnet accrues fluence we replace the constant terms in equation (4) with integration across temperature and introduce time-varying parameters,

$$\int_{T_{operation}}^{T_{critical}(t)} w(T, t)\, dT = \int_{T_{operation} + T_{margin}(t)}^{T_{critical}(t)} c_p(T)\, dT \qquad (5)$$

where $c_p(T)$ is the specific heat capacity of the stabilizer material and $w(T, t)$ is the Winger energy release (recovery) profile of the stabilizer material (example left axis Figure 4). $T_{operation}$ is the operating temperature of the fusion magnet, $T_{margin}$ is the minimum temperature disturbance from operation that results in a quench, and $T_{critical}(t)$ is the temperature at which the quench initiates. In this equation, the Wigner energy profile and critical temperature are recognized to vary with time, resulting in a varying temperature margin with time. Equation (5) can be rearranged to more clearly describe the dynamics of temperature margin,

$$\int_{T_{operation}}^{T_{critical}(t)} c_p(T) - w(T, t) \, dT = \int_{T_{operation}}^{T_{operation}+T_{margin}(t)} c_p(T) \, dT \tag{6}$$

where $T_{margin}$ must clearly reduce as $w(T, t)$ increases with fluence. Equation (6) also shows how $T_{margin}$ reduces as $T_{critical}(t)$ reduces.

Equation (6) describes how the Wigner release grows with time, effectively reducing the specific heat capacity of the magnet's stabilizer material due to the storage of energy from defect production. The stored energy provides additional energy to induce a quench following an external energy input. The definition described by equation (6) was developed with a few assumptions,

1) Linear relationship between time and fluence. This assumption is true for steady-state constant fusion power output, which is the ideal operational mode. In reality, transient states of operation would mean that energy storage rates in the stabilizer material would change with time due to changes in neutron flux, and must be accounted for during operation (e.g., in digital twins).

2) The Frenkel pair defect recombination time constant is much less than that of the temperature dynamics resulting from a thermal energy disturbance. The justification of this assumption is explored in Section III of this document.

Equation (6) can be further simplified by assuming that $T_{critical}$ does not change with time and then integrating the temperature-dependant term now bounded by constant limits,

$$T_{margin} = \left( 1 - \frac{W(t)_{T_{operation}}^{T_{critical}}}{C_{T_{margin}}^{T_{critical}}} \right) (T_{critical} - T_{operation}) \tag{7}$$

$W(t)_{T_{operation}}^{T_{critical}}$ is the Wigner energy released upon heating from operating temperature to critical temperature (grows with time) and $C_{T_{margin}}^{T_{critical}}$ is the amount of energy required to heat from operating temperature plus temperature margin to the critical temperature of the magnet.

While equation (7) is not entirely physically accurate due to the simplifying assumption (because superconductor performance degrades with fluence) equation (7) provides some insight into the relationship between the physical properties of the material and the performance of the magnet. However, if the actual Wigner energy storage rate in stabilizer materials is much greater than the rate of degradation of superconductors, then equation (7) could be used to predict the life of a superconducting fusion magnet. With the sparse experimental data currently available, right axis Figure 6, it is not possible to say if this equation is an accurate predictor of fusion magnet life.

### C. The Well's Number

Equation (6) provides a more general and robust description of temperature margin than was previously described by equation (2). Time t=0 marks the beginning of reactor operation. $w(T, t)$ equals zero at t=0. Equation (6) then collapses into equation (2). As t increases, $T_{margin}$ approaches zero and can become negative. t=t$_{critical}$ describes another special condition where $T_{margin}$ equals zero. t$_{critical}$ is first described in this work and is termed "critical span".

Time resets (t=0) following a maintenance temperature cycle of the magnet. A maintenance temperature cycle is a planned heating cycle of the magnet to anneal defects and release stored energy, resulting in a recovery of $T_{margin}(t)$.

In other words, the critical span is defined as: "Continuous operating duration until the temperature margin of a fusion magnet is effectively zero and a magnet quench becomes potentially spontaneous."

Time $t$ increases linearly with continuous operating duration between maintenance temperature cycles of the magnet, with the assumption of linearity between time and fluence. A maintenance temperature cycle permits the recombination of defects responsible for the reduction in $T_{margin}$. The temperature cycle can be to room temperature, but heating to lower temperatures, such as 100K, would anneal defects sufficiently to recover $T_{margin}(t)$ and reset t to zero.

Reductions in $T_{critical}(t)$ due to irradiation damage to the HTS are more difficult to recover than reductions in $T_{margin}(t)$ due to damage to the stabilizer material [4], [35]. Superconductor irradiation damage will likely dictate the ultimate lifetime of the magnet while stabilizer irradiation damage will be an essential consideration in the continuous operating regimes of the machine.

A non-dimensional number defined by the log of the ratio between the fluence at critical span, and the continuous operating fluence accrued since the last maintenance temperature cycle enables interpretation of the quench risk for a superconducting fusion magnet,

$$We = log \left( \frac{\Phi_{Critical}}{\Phi_{Continuous}} \right) \tag{8}$$

This non-dimensional log ratio is defined in this work and is termed the Well's number. The Well's number is defined by the log ratio of critical fluence, $\Phi_{Critical}$, to continuous operating fluence, $\Phi_{Continuous}$. The standard, full power year (FPY), is typically used to linearize between fluence and time, it is suggested to adhere to this practice and describe $t_{Critical}$ and $t_{Continuous}$ in terms of FPY,

$$t_{Critical} = \frac{\Phi_{Critical}}{\phi_{FPY}} \tag{9}$$

$\phi_{FPY}$ being the continuous steady-state flux necessary to produce a full power year of energy. The assumption of linearity between fluence and operating time only applies during design. During digital twin simulations, or reactor operation, fluence is a state that is either observed or measured and the complexities of the relationship between fluence and time can be accounted for. During digital twin simulations or operation of the reactor, equation (9) is no longer valid.

As the Well's number decreases to zero, the risk of a quench increases. Once negative, the fusion magnet contains enough stored energy to spontaneously increase to the critical temperature and initiate a quench without a temperature fluctuation. A non-linear increase in the likelihood and severity of fusion magnet quenches is expected with increasing fluence due to the combined effects of resistivity change, thermal conductivity change, and stored energy release. Future iterations of the Well's number may include these parameters in this interpretation of risk. Normalizing against critical fluence provides an even comparison between design differences such as neutron spectra at the magnet or stabilizer material.

The permissible Well's number for a fusion magnet design is a combination of numerous factors. The performance of the quench detection and control system is measured by the time it takes for the system to detect and halt a quench. This system will dictate how much material property change in the magnet will be permissible. As material properties change with defect production, quenches become more severe. The relationship between material property changes and the performance of the quench detection and control system will dictate what fraction of the critical span the magnet can safely operate to.

To extend the safe continuous operating duration of a magnet, an engineer can either increase the critical fluence or improve the quench detection and control system.

### D. Energy Balance

The Wigner energy release is given engineering context by considering specific/measured irradiation conditions. The left-hand side of a hypothesized equation describes the energy released due to defect recombination between a temperature range. The right-hand side describes the associated temperature increase of the material after absorbing the released thermal energy,

$$\int_{T_{operation}}^{T_{critical}} w_{Losehand}(T)\, dT = \int_{T_{operation}}^{T_{final}} c_p(T)\, dT \qquad (10)$$

when heated from $T_{operation}$ = 20K to $T_{critical}$ = 40K, after exposure to the Losehand et al. [9] irradiation, with $w_{Losehand}(T)$ designating the Wigner energy release profile measured during the experiment (left axis Figure 4). Cu releases enough energy to heat from $T_{operation}$=20K to $T_{final}$=29.6K. A piece of Cu can not simultaneously heat from 20K to 40K and 20K to 29.6K, so this calculation is not a physical representation of reality and it does not predict temperature margin, equation (6) is required for this. This calculation shows that the magnitude of energy released by heating between $T_{operation}$ and $T_{critical}$ is of practical significance.

### E. Temperature Margin Reduction with Fluence

Figure 9 and Figure 10 show calculations of temperature margin using equation (7) and equation (6) respectively. Together these figures describe the possible consequences of the stored energy release and provide a comparison of these consequences to the reduction in performance of HTS superconductors.

The operating temperature and initial critical temperature were taken to be 20K and 40K for both calculations, respectively. Specific heat capacity was referenced in Simon et al. [26]. The calculation in Figure 9, performed with equation (7), assumes critical temperature does not change with fluence. The calculation in Figure 10, performed with equation (6) assumes critical temperature degrades with fluence due to damage to the high-temperature superconductor [3]. The Wigner energy release profile in Losehand et al. [9] was normalized by total energy release and fluence. The normalized profile was scaled by measured energy release and fluence values reported in the third and fourth columns of Table III. These extrapolations provide estimates of possible Wigner energy release profiles for various irradiation conditions. These extrapolations are only expected to be relevant up to a fluence of $3.5 \times 10^{18}$ n/cm$^2$. After $3.5 \times 10^{18}$ n/cm$^2$ in Figure 5 the linear trend between resistivity and fluence dies and a higher order correlation would be required to continue the extrapolation. Also, the data describing HTS degradation used in the calculation of Figure 10 is only measured up to $3.5 \times 10^{18}$ n/cm$^2$ [3].

*1) Accounting for Wigner energy release only:* The furthest left profile in Figure 9 is produced from a measurement following a low-fluence 11 MeV deuteron flux. This flux exhibits low defect storage efficiency but high primary knock-on energy resulting in a high energy stored per particle relative to the other experiments, see Table III. The fluence of the experiment is much lower than the extrapolated fluence values used in the analysis. These lines should be interpreted with the understanding that there is likely great error in this estimate due to the distance of the linear extrapolation.

The black, green, light blue, blue, and red solid lines are produced from measurements following various neutron irradiations. The difference between the green, light blue, blue, and red solid lines is likely a result of the difference in primary knock-on energy due to differences in neutron spectrum energy. The difference between the black solid line and the other solid lines can be speculated and may be a result of how the measurements were taken or differences in spectra.

The results used in the calculation of the black line reported the highest energy stored out of all the reported studies in Table III. Also, this study exhibits the highest stored energy per incident neutron, implying that the stored energy measurement may be artificially high due to measurement error. However, the stored energy per incident particle is less than that reported by the 11 MeV deuteron irradiations, which is physically realistic.

It is noteworthy that the fluence for the Losehand et al. study was not reported and was estimated in this work using the resistivity change to fluence correlation measured in Nakagawa et al. [10], as discussed in section II. The error in this estimation is low because both experiments were performed with the same spectra and likely the same equipment.

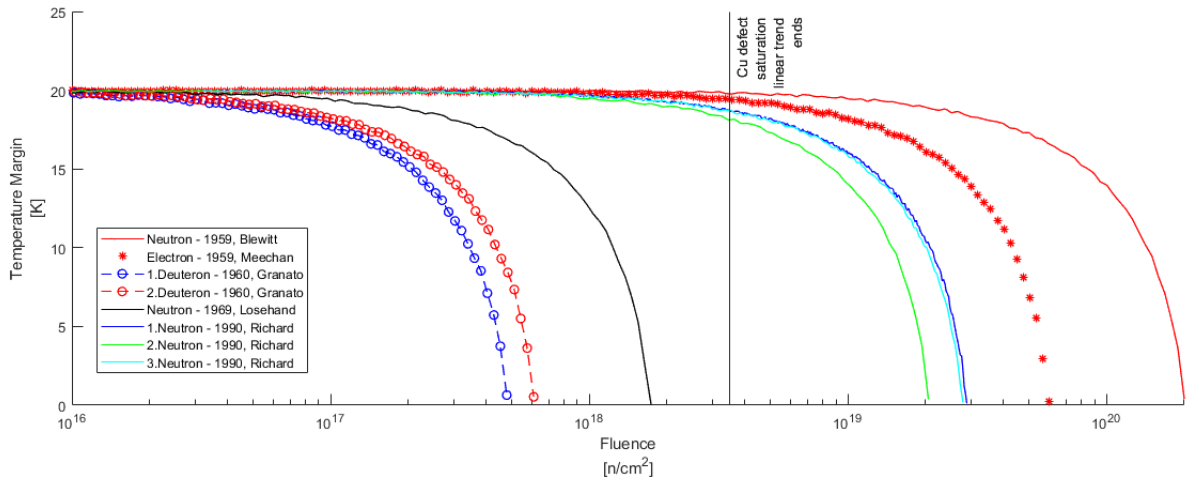

Fig. 9: Temperature margin reduction with fluence resulting from Wigner energy release in the magnet's stabilizer material. These predictions are produced using an equation derived from first principles, equation (7). This equation is a special case of equation (6) where the critical temperature is treated as constant. The difference between each line is due to the difference in how the stored energy release was predicted for each case. Linear extrapolations between zero fluence and the reported fluence from experiments referenced in Table III were used to predict stored energy release for each case. The linear extrapolation was assumed to be valid up to $3.5 \times 10^{18}$ n/cm$^2$ given the linear trend of Cu resistivity change with fluence observed in Figure 5.

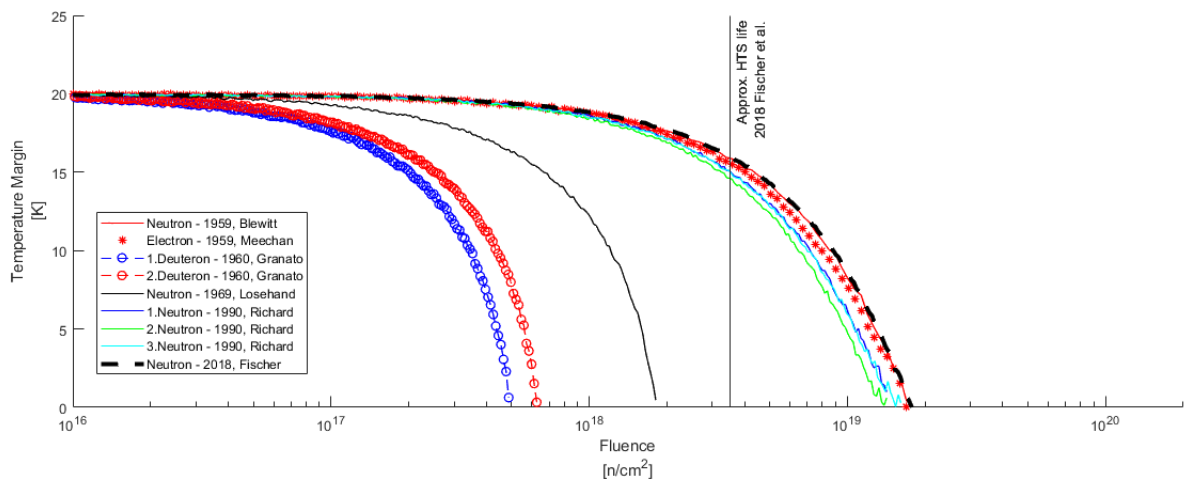

Fig. 10: Temperature margin reduction with fluence resulting from Wigner energy release and HTS degradation. These predictions are produced using an equation derived from first principles, equation (6). This is the same calculation presented in Figure 9 but with the consideration of a reducing upper limit on the integrals with fluence, $T_{critical}(t)$, in equation (6) to account for HTS degradation. Fluence is assumed to be linearly proportional to time for steady-state operation and the $T_{critical}(t)$ term is assumed to reduce with fluence as described by Fischer et al. [3]. Values are only reported up to $3.5 \times 10^{18}$ n/cm$^2$, so these predictions should only be considered valid up to $3.5 \times 10^{18}$ n/cm$^2$. Comparison of Figure 9 and Figure 10 describes that energy storage in the magnet's stabilizer material is an important consideration in the continuous operation of the machine, and may even play a rate-determining factor role for continuous operating regimes.

The profiles in Figure 9 depict the reducing resilience of the superconducting fusion magnet against temperature disturbance as the magnet's fluence increases. The critical fluence, shown in equation (8), for each profile is the x-intercept. Critical fluences range from $5 \times 10^{17}$ n/cm$^2$ to $2 \times 10^{20}$ n/cm$^2$ for the considered irradiation conditions. $1.74 \times 10^{18}$ n/cm$^2$ and

$2.85 \times 10^{19}$ n/cm$^2$ are taken to be the minimum and maximum critical fluence estimates for magnet systems controlling fusion reactions.

The trend of red stars is produced using data from an electron irradiation experiment. This spectrum exhibits high storage efficiency but low primary knock-on energy, storing predominately Frenkel pairs. The actual recovery profile, $w(T, t)$, for this case, would be different than that used in this work. Defects produced from this spectra (Frenkel pairs) anneal at 45K. The actual profile for this experiment would exhibit more annealing at lower temperatures and the curve of red stars, in both Figure 9 and Figure 10 would shift to the left. However, it is not physically realistic to irradiate fusion magnets with a flux of electrons.

*2) Accounting for Wigner energy release and HTS degradation:* Figure 10 is the same calculation as presented in Figure 9, but includes the reduction in critical temperature of HTS reported in Fischer et al. [3] due to collisions that occur during neutron irradiation. The thick dotted black line depicts the upper limit on magnet life set by the degradation of the HTS. The life of the magnet is determined as a combined effect of HTS and stabilizer irradiation damage. It is important to understand the behaviour of stored energy release due to annealing in Cu with greater precision to conclude the safe continuous operating duration for fusion magnet systems.

The next steps are to produce more precise recovery predictions of magnet stabilizer materials for the conditions of fusion magnets and relate them to the degradation rate of high-temperature superconductors. Temperature cycles will likely be necessary to mitigate quench risk. Fortunately, cycles of irradiation, annealing, and irradiation do not cause an accumulation or embrittlement of pure Cu [36]. This supports the claim that the life of the magnet will be limited by HTS performance degradation.

Estimated HTS degradation, black dotted line in Figure 10, puts an estimated upper-limit on magnet life at a fluence of $1.77 \times 10^{19}$ n/cm$^2$. Measurements of the critical temperature of HTS have only been performed to $3.5 \times 10^{18}$ n/cm$^2$. The critical fluence due to HTS degradation alone may be lower than the estimated $1.77 \times 10^{19}$ n/cm$^2$, more experiments are necessary. Critical fluence reductions due to irradiation damage to the HTS can be recovered, but some damage is permanent [4]. Annealing in an Oxygen atmosphere has been shown to encourage this recovery [35]. Whereas, changes due to defect production in Cu can be fully recovered.

*3) Extending plant operating duration:* Annealing dynamics between 20K and 40K change significantly between materials. Au exhibits very little recovery occurring between 20K and 40K [11]. Additions of Au to Cu reduce the heat released between 20K and 40K due to reduced interstitial migration [37]. Piani & Aspeling examined the recovery behaviour following fast neutron and thermal neutron irradiation of pure Cu and a Cu-0.01Au at.% mixture [37]. The study irradiated these samples at 4.4K, performed an annealing cycle between 30K and 400K, and examined the percent recovery occurring between temperature increments. The temperature increment of interest in this work occurs between 30K and 60K. The list below examines the percent recovery for Cu and Cu-Au samples following 4.4K irradiation of both thermal and fast neutron irradiation for a heating range of 30K and 60K.

- Cu sample and thermal neutrons: recovered 55.8% of its resistivity following a thermal neutron irradiation resulting in a resistivity change of 0.025 nΩm.
- Cu sample and fast neutrons: recovered 32.6% of its resistivity following a fast neutron irradiation resulting in a resistivity change of 0.250 nΩm.

- Cu-Au sample and thermal neutrons: recovered 41.4% of its resistivity following a thermal neutron irradiation resulting in a resistivity change of 0.030 nΩm.
- Cu-Au sample and fast neutrons: recovered 28.3% of its resistivity following a fast neutron irradiation resulting in a resistivity change of 0.267 nΩm.

The magnet stabilizer material's composition can be engineered to enhance the maximum continuous operating duration of the fusion power plant following a deeper understanding of recovery behaviour in magnet stabilizer material and its relationship to plant performance.

## VI. Conclusion

A superconducting fusion magnet can carry circa 100 kA and contain stored electromagnetic energy at the scale of gigajoules. When superconducting, the material exhibits zero resistance to current. Irradiation damage to superconducting material degrades critical current, which defines the current density when superconductivity is lost. After losing superconductivity, in a magnet leveraging HTS, current is favorably carried by the stabilizer material instead of the HTS. The stabilizing material is typically cryogenic Cu and exhibits heating according to Joules law when carrying current. The transition of a superconducting magnet to normal state during operation is called a quench. Due to the scale of electromagnetic energy contained in the magnet during operation, quenches can be devastatingly destructive. Irradiation damage to the stabilizer material reduces the material's thermal conductivity, specific heat capacity, and increases normal state resistivity. These changes occur due to energy stored as defects in the crystalline lattice of the stabilizer material. These changes induce the following engineering challenges:

- Greater difficulty in ensuring a zero derivative temperature gradient across the cross-section of the magnet due to reduced thermal conductivity.
- Increases in the amplitude of temperature perturbations due to the release of energy when interstitial defects become mobile and anneal. This energy release is effectively observed as a reduction in specific heat capacity.
- Higher heat loads during a quench due to increases in resistivity to the current carrying material during this failure mode.

Fluence to the magnet increases the likelihood and aggressiveness of quenches. Planned heating cycles should be considered in fusion reactor designs which comprise superconducting magnets. The temperature cycle mitigates the risk described in this paper by safely annealing defects in the stabilizer material. Avoiding the application of superconducting material in fusion magnets may be practical until the performance of superconducting magnets under neutron irradiation is characterized.

This work examines experimental measurements of resistivity increases and energy released from Cu following cryogenic irradiation. The data is extrapolated across fluence for predictions of fusion magnet performance following irradiation. Data at higher fluence values is necessary for more precise predictions. The limits of the application of these extrapolations is defined by observed saturation phenomena for resistivity change.

A new definition for temperature margin, the temperature difference between operation and quench runaway, was provided that accounts for some of the material property changes to the stabilizer material and accounts for the degradation of critical current in the HTS with irradiation damage. A non-dimensional number, termed the Wells number, is described to interpret the risk of a quench after accruing fluence to the fusion magnet. In its current form, the Wells number normalizes for changes to the specific heat capacity of the stabilizer and degradation of the critical temperature of the HTS. Future iterations of the Wells number may be more comprehensive.

The temperature margin was found to reduce with fluence. Temperature margin predictions were performed including only stabilizer material property changes and including the combination of stabilizer material property changes with the reduction in critical current of HTS. It was found that both HTS degradation and stabilizer material property changes are important considerations in this calculation. However, this calculation did not include temperature distribution or quench temperature dynamics calculations. Both of these calculations would show non-design-favorable change with stabilizer irradiation damage. This observation increases the likelihood that the risk due to stabilizer irradiation damage is greater than that quantified in this paper.

Temperature distribution and quench dynamics calculations that account for material property changes due to irradiation damage are the next step in this work. Experiments to precisely measure the recovery behaviour of stabilizer materials following irradiation would contribute meaningfully to mitigating this risk. Ideally, in-situ fusion magnet quench trials should be performed in a neutron flux. Mitigating the risk of fusion magnet quenches is a necessary step toward specifying the application of superconducting material in fusion power plants.

## ACKNOWLEDGMENTS

This work has been funded by the EPSRC Energy Programme [grant number EP/W006839/1]. To obtain further information on the data and models underlying this paper please contact PublicationsManager@ukaea.uk.

Thank you to Charles Hirst, Max Boelininger, and Luke Hewitt for the interesting technical discussions we shared.